\documentclass[aps,prd,nofootinbib,twocolumn,longbibliography]{revtex4-1}
\usepackage{graphicx,amsmath,amssymb,bbm}
\usepackage[colorlinks, citecolor=blue, linkcolor=blue]{hyperref}

\newcommand{\be}{\begin{equation}}
\newcommand{\ee}{\end{equation}}

\begin{document}
\title{ \Large A critical look at strings}
     \author{Carlo Rovelli}
     \affiliation{Centre de Physique Th\'eorique de Luminy\footnote{Unit\'e mixte de recherche du CNRS et des Universit\'es de Aix-Marseille I, Aix-Marseille II et Toulon-Var; affili\'e \`a la FRUMAM.}, Case 907, F-13288 Marseille, EU}
\date{\small  \today}
\begin{abstract}
\noindent 
Following the invitation of the editors of \emph{Foundations of Physics}, I give here a personal assessment of string theory, from the point of view of an outsider, and I compare it with the theory, methods, and expectations of my own field.

\end{abstract}
\maketitle

\section{Introduction}

\rightline{\begin{minipage}{7.6cm}
{\em ``Perch\'e i discorsi nostri hanno a essere intorno
al mondo sensibile, e non sopra un mondo di carta."}\\
Galileo Galilei, {\em Dialogo sopra i massimi sistemi}
\end{minipage}}
\vspace{4mm}
 
 \noindent
I am not an expert on strings. I follow the results announced as main string achievements, but I have not worked in the field. I have therefore much hesitated before accepting the invitation by \emph{Foundations of Physics} to express a view on the theory.  I have eventually decided to accept, in the hope of giving a contribution to the overall debate on the theory, because a central problem addresses by string theory is also addressed by the research direction in which I work. 

For a considerable number of years, strings have represented a huge intellectual investment, aiming at a complete theory capable to describe the world at the elementary level, including quantum gravity.  Today, the problem is obviously not yet solved.   String theory is incomplete, far from describing precisely our real world, and its foundation is poorly understood. But the difficulties of a similar task are arduous and advances are necessarily slow. Strings provide tantalizing hints, partial answers, intriguing mathematical tools, and the tentative architecture of a grand overall picture to solve the problem.  

In such a situation it is hard to evaluate string theory in isolation.  An evaluation can only be made by comparing the theory with alternative research directions.  Following the indications of the editors of \emph{Foundations of Physics}, I try here to asses the results of string theory by comparing them with results and methods of my own field of research.  I hope that this can contribute to put the results of string theory in perspective, and seeking a sober evaluation of the relative merits and the relative potential of different research directions, in a field where the final answer is not yet known. 

String theory has recently been evolving into a toolbox, with tentative applications to fields such as QCD, strongly interacting fluids, or pure mathematics. I will not comment on the interest of string theory for these fields. This should be evaluated by QCD theorists, condensed matter physicists, or mathematicians. I focus on the motivating claim of string theory, which is to describe the real world beyond what is well accounted-for by the particle-physics standard-model and classical general relativity, and in particular to concretely describe the regimes where the quantum property of gravity cannot be neglected. This last problem is my own specific field of interest.  

\section{What we know, what we do not know, and what is the problem}

Before addressing merits and shortcomings of the tentative \emph{solution} provided by string theory, let me briefly recall the \emph{problem} on the table, which string theory means to solve.   The Standard Model and classical general relativity are spectacular theories that have enjoyed an empirical success with few --if any-- equal in the history of science. Today, these theories (with neutrino mass and cosmological constant) seem able to account for virtually anything we can measure, with the notable exception of  dark-matter phenomenology. These are the currently established fundamental theories that summarize what we know about the physical world at the most elementary level  we can access.  But this set of theories does not allow us to compute what happens in all physical regimes, is patchy,  and manifestly incomplete. 

Specifically, if we want to compute the scattering amplitude of two point-particles interacting gravitationally, as a function of their center of mass energy $E$ and impact parameter $b$, we can do so using (non-renormalizable)  perturbative quantum general relativity as an effective theory, but predictivity breaks down when $E$ becomes of the order of $b$ in Planck units.  Therefore current established theory gives no prediction whatsoever on what happens to particles that scatter at that energy.   Such lack of predictivity is particularly relevant in important physical situations, such as early cosmology, some aspects of black hole physics, and our understanding of the short-scale structure of physical space. We need new theory to understand this physics.

Furthermore, there are major theoretical and conceptual shortcoming of the current theory.  Ultraviolet divergences appear to indicate that there is something important we miss at short scale.  General relativity is a beautiful theory with a tight formal structure and a minimal number of free parameters, but the Standard Model is a patchwork with a number of free parameters that calls for an explanation.  The conceptual structure of the Standard Model includes aspects (fixed spacetime, global Poincar\'e invariance, local energy conservation...) which play a major role in dealing with its quantum properties, but are profoundly different from that of general relativity (dynamical spacetime, no global Poincar\'e invariance, no local energy conservation, general covariance...): if we want a coherent picture, we need a way to combine the two.  In particular, if the central physical tenet of general relativity is correct, namely if the geometry of physical space is a physical field, then the quantum character of this field imply that physical geometry is ``quantum geometry". What is quantum geometry?  Can we find a complete and consistent theoretical picture where these issues are resolved?  

There are \emph{two} problems raised by this situation. The first is to {complete} the picture and make it {consistent}.  This is the problem of \emph{quantum gravity}, since what is clearly missing is the understanding of quantum gravitational physics.  A second, distinct, problem, is \emph{unification}, namely the hope of reducing the full phenomenology to the manifestation of a single entity.  (QCD completes the standard model and is consistent with  electroweak theory, but is not unified with it.)  String theory is an attempt to solve the two problems at once, namely to provide a quantum theory of gravity \emph{within} a unified picture (a ``final theory").  

Below, I discuss the extent to which these problems are solved by the current state of string theory, in comparison with other approaches.  I focus on the approach to quantum gravity in which I work, loop quantum gravity. (Recent reviews of string theory to which I am particularly indebted are \cite{Mukhi:2011zz}, and \cite{Blau:1900zza} which focus on strings as a theory of quantum gravity. A recent overview of loop gravity, with relevant references is \cite{Rovelli:2010bf}, a technical introduction is in \cite{Rovelli:2011eq}.)  It is important to stress upfront, however, that the problems addressed by strings and loops do not coincide. Both theories aim at a quantum theory of gravity in order to complete the current theoretical description of the world and and make it coherent, but string theory assumes the working hypothesis that this can only be achieved in the context of a unified theory, capable of addressing also questions that are outside the scope of loop gravity (such as: why these matter couplings and not others?  why four dimensions? what is the final theory of nature? and so on.)

\section{Ultraviolet finiteness}

A major achievement of string theory is the control of the ultraviolet divergences of conventional quantum field theory.  An actual proof of ultraviolet finiteness is still lacking. At least, I have searched and much asked around, but I have not yet been able to find a reference with such a proof.  But string theorists appear to be convinced of finiteness, and I believe them.  

Indeed string theory provides intuitive ways for seeing how singularities are resolved. When point-particles scatter at very high energy, the stringy degrees of freedom ``open-up", effectively spreading over a finite spacetime region, smoothing out the interaction region.  From a different perspective, in the perturbation expansion the Riemann surfaces corresponding to high momentum are the ``thin" ones, but a modular transformation relates such surfaces to non-degenerate ones, effectively avoiding the ultraviolet regime.  The reason for the stringy  ultraviolet finiteness can therefore be traced to the very hypothesis at the basis of string theory, namely the existence of an infinite number of degrees of freedom besides the ones we see, defining extended elementary objects. These can smooth out the standard quantum field theory divergences. 

These result can be compared with the ultraviolet finiteness of loop quantum gravity.  Here the proof of ultraviolet finiteness is straightforward.\footnote{Infrared finiteness can be proven as well in an appropriate version of theory. See the review articles quoted above for specific references.}  Unlike string theory, the physical ground for the ultraviolet finiteness of loop quantum gravity does not require any additional input beside general relativity and quantum mechanics, and it can be understood as follows.  Since the geometry of space is a quantum field, it has quantum properties. In particular, the spectrum of the geometrical quantities can be computed and turns out to be discrete. At short scale, spacetime geometry is therefore effectively discrete, in very much the same manner in which the energy of a harmonic oscillator is discrete.  Therefore there is literally ``no room" for ultraviolet divergences in the theory: there is no short-distance beyond the Planck scale. This becomes manifest in computing transition amplitudes, where one sees explicitly that high momenta are cut off at the Planck scale.\footnote{Geometry discreteness does not break Lorentz invariance in loop gravity because it is quantum mechanical. Eigenvalues do not transform continuously under a continuous symmetry. (The $\hbar/2$ eigenvalue of the angular-momentum component $L_z$ does not transform continuously under rotation.) A boosted observer does not measure Lorentz-contracted discrete lengths: he measures a continuously-deformed probability distribution for the \emph{same} spectrum of lengths. The hard-to-die idea that the existence of a minimal length at the Planck scale necessarily breaks Lorentz invariance is plainly wrong.} 

Ultraviolet finiteness is therefore a main achievement both for string theory and for loop gravity. But it is realized differently in the two theories.   In both cases, it reflects an intrinsic physical limitation in measuring distances shorter than the Plank scale. But in the first, it follows from a novel hypothesis about Nature; in the second, it is a direct consequence of quantum theory and general relativity. 

\section{Quantum geometry}  

The peculiar features of general relativity, and in particular its large gauge symmetry, have followed a curious fate along the evolution of string theory.  To start with, general relativity was treated as a conventional field theory in the string context. It was considered the effective low-energy manifestation of something else, like Fermi weak-interactions theory,  and dealt with using basic quantum field theoretical ``non-general-relativistic" tools: expanding around a fixed background, relying heavily on global Poincar\'e invariance and so on. That is, relying on notions that are at odds with the symmetry of general relativity.  In the hands of theoreticians mostly coming from the particle-physics tradition, general relativity was treated in a way that appeared to general relativists to betray its central physical ideas.  But the peculiar features of general relativity and of the quantum aspects of spacetime have eventually resurfaced and are playing an increasingly important role in string theory today. 

The main effect of the large symmetry group of gravity, and the main teaching of general relativity, indeed, is that the world is not a given spacetime over which dynamical degrees of freedom evolve.  Rather, spacetime itself is a dynamical entity.  In a quantum theory, spacetime itself is a quantum entity, whose structure cannot be assigned a priori.   A  number of  developments of string theory aim at coming to terms with this deeply unconventional and novel aspects of the world that is directly implied by the physics of general relativity and quantum theory. For instance, with states that have no natural continuum spacetime description (say, a vacuum which is a tensor product of conformal field theories).  Similarly, the difficulties of defining local bulk observables in a general relativist context which have  long been discussed in the quantum gravity literature are now being increasingly discussed in the string literature.

Of course, much physics can be derived by choosing a background and computing around it.  A state of a background-independent theory constitutes a background, and physics around that state will be -obviously- background dependent. There is no conflict between this background dependence about a chosen (``vacuum") state, and  fundamental background independence, any more than there would be between quantum electrodynamics and its expansion about the field of an atomic nucleus. The difficult problem is another one: whether the full definition of the theory, and in particular the characterization of its degrees of freedom, requires a background to start with or not. For instance, if we reinterpret  general relativity as the theory of small fluctuations around a fixed space-time, we lose most of the interesting phenomena it predicts, such as the Schwarzschild solution, the dynamics of the universe, black holes's horizons, and so on. In other words, the problem is whether or not we have a quantum theory with a clear definition of a state space capable of listing all possible background states. 

There have been numerous beautiful attempts to find this fully background independent formulation of string theory, such as string field theory, matrix theory, holography... But  full background independence of string theory is not yet properly understood.  

The way this fundamental issue is addressed in string theory is often indirect.  For instance, attempts are made to describe the bulk  quantum geometry of spacetime by using the ADS-CFT conjecture, thus trying to describe what we do not know (quantum gravity) in terms of conceptual tools that we control (flat-space quantum field theory on the boundary).  Analogously, the string theory calculations of black hole entropy exploit the relation between the strong-coupling genuinely-gravitational regime of interest, and the weak-coupling regime where conventional flat-space tools can be used, and states can be counted.  Again, string cosmology often addresses the highly non-Minkowskian geometry of early cosmology by an hypothesis, that sounds bizarre to relativists: an overall larger Minkowski space where everything happens.  

In all these cases, instead of addressing the real problem, which is to learn how to do physics where background spacetime plays no role, the strategy is to try to circumvent the problem, bringing back the calculations to the familiar pre-general-relativistic conceptual framework. The reason of this, of course, is not lack of imagination or courage from string theorists.  String theory gives glimpses and hints of how a genuine theory of quantum geometry could be, with general states having no Riemannian spacetime interpretation at all ---like a general state of a quantum particle is not necessarily similar to a classical localized particle--- but for the moment it is far from providing a complete  coherent picture of quantum geometry. 

This must be compared with the picture of quantum geometry offered by loop gravity.  Contrary to the string case, loop gravity addressed upfront  the problem of describing the fundamental degrees of freedom of a theory without a fixed background spacetime. The result is that everything is conceptually clear, fully general relativistic, and well defined.  There is a Hilbert space, whose states have a clean interpretation as quantum states of the geometry. These do not live over a background, but themselves build-up spacetime.  The quanta of the theory are ``quanta of space", quantum bricks that build up spacetime.   The mathematics of quantum geometry is clear at the level of mathematical physics, as well as at the conceptual physical level.  A formalism for computing well defined background-independent observables, as well as perturbing around a given background, is known.  These techniques are perhaps unfamiliar to many, and might look strange at first sight, but so did string theory for many years, before becoming fashionable. 

It seems to me that the clarity of the picture of quantum geometry is definitely a plus for loop quantum gravity that lacks in string theory.

\section{Overall picture}

The beauty of string theory, on the other hand, is that it offers a tentative overall picture capable of bringing together in a natural and compelling way so many aspects of the world.  It provides an ultraviolet consistent theory of gravity and at the same time has natural room for gauge symmetry,
unification, holography, all fused in an interrelated net that suggest the existence of a compelling overall architecture. Even not working in the field, one cannot fail to appreciate the tantalizing aspects of the relations unraveled by the string research.  It is very tempting to believe that beyond all these relations there should be a remarkable coherent edifice.

The difficulty is that for the moment we see only bits and pieces of the hypothetical complete edifice.  In particular, we do not see the foundations: the basic degrees of freedom and the basic equations. The sentiment that this beautiful underlying theory \emph{should} exist is strong among the people immersed in string theory, and is reinforced  by the discovery of the beautiful relations --dualities-- relating the different bits.  It is difficult for an outsider 
to fully appreciate the support of this sentiment, but string theorists appear to be convinced of the existence of the  underlying theory.   They might be right, but the fundamental theory, if it exists, is still outside our  control. Until we see it, its beauty and its physical consistency are hypothetical. 

This can be compared with loop quantum gravity. The scope of the theory is much narrower, because the theory does not pretend to be a unified theory, does not select the matter couplings, and does not aim at being the final theory of the world.  But the elementary degrees of freedom, which are the quanta of space, or, equivalently, the quanta of gravitational field, are clearly defined.  The basic operators are well defined.   The dynamics can be compactly presented with three equations. The overall structure of the theory is complete and simple.

Loops and strings differ in another key respect.  Strings are based on a definite physical hypothesis: elementary constituents of the world are  extended objects.  The hypothesis might be right. Or wrong. The world might not be supersymmetric and 10 dimensional.  

Loop gravity, on the other hand, is grounded in quantum theory and in the symmetry underlying general relativity, a symmetry today generally expected to survive at high energy. Loop gravity is just a \emph{general covariant} quantum field theory, with degrees of freedom reducing to Riemannian geometry at low energy.   Loop gravity can very well turn out to be wrong as well, of course.  But if  the theory is wrong, it must be so for some more subtle reason, which, in any case, would still teach us something about the quantum world at the general covariant quantum level. 

 Of course, assuming that the basic physical tenets of general relativity and quantum theory remains valid at the Planck scale is an extrapolation.  But extrapolation has always been the most spectacularly effective tool in science. Maxwell equations, found in a lab, work from the atomic to the galactic scale. Up to contrary empirical indications, always possible, a good bet is that what we have learned may continue to hold.  
 
\section{Describing \emph{this} world}

Let me now come to what I see as a serious shortcoming of string theory.  The interest in the theory exploded around 1985, when $E8\times E8$ and the heterotic string appeared to be the \emph{unique} viable option, and the low-energy field theory of such a string, compactified to 4 dimensions on a suitable class of 6-dimensional manifolds, was shown in a classic paper of Candelas, Horowitz, Strominger and Witten \cite{Candelas:1985en} to yield qualitatively correct phenomenological properties, including parity violation. The central promise of that paper and the hope it raised was that a realistic string theory incorporating and generalizing the Standard Model, plus gravity, was round the corner. I think that one can safely say today, in hindsight and despite the defining historical role played by the paper, that the hope grounding that paper, which sparked all that interest, was misplaced. 

String theory would be in a stronger position, if it could exhibit a mechanics  yielding the $SU(3)\times SU(2)\times U(1)$ gauge group, the particle content of our world, the three generations, no supersymmetry at our scale and so on. Understanding \emph{this} was its original aim.  So far, string theory fails to describe our world as see it. It describes, instead, lots of worlds, in all sort of higher dimensions, generally with cosmological constant having the wrong sign, with ``microscopical" internal spaces of cosmological size, and so on. This is a beautiful theoretical world, with marvels and surprises, but where is \emph{our} world in it? Until the description of our world is found in this immense paper edifice, it seems to me that caution should be maintained. 

This can be compared with the situation in loop  gravity, or with other approaches to this problem, that might shed light on some of these issues, like for instance Alain Connes's non commutative geometrization of the Standard Model \cite{Chamseddine:1991qh}.  Again, loop gravity does not pretend to provide a unified picture of nature, to tell us what is the matter content of the universe, or to determine the number of dimensions of spacetime.  But the theory is compatible with a description of the world as we see around us: four dimensions, no supersymmetry, fermions and a  certain Yang-Mills gauge group.  Like all successful physical theories developed so far (QED, QCD, or general relativity) it is compatible with unphysical couplings.  The ambition of loop gravity is not to solve all problems of physics and provide the final theory. Its ambition is to provide a consistent theory of quantum gravitational phenomena,  coupled with the matter that we find (with experiments) in the world.

On the other hand, scattering calculation around the Minkowski background in loop quantum gravity are being developed, but they are in a far more primitive stage than the scattering calculations one can do with string theory.  Also, in the last couple of years, loop gravity has seen the development of an explicit formulation of the theory that includes fermions and Yang-Mills fields, and of a technique to compute scattering amplitudes around the Minkowski background.  But these developments are recent, and the results are preliminary.

\section{Unification}

There is one issue in which string theory appears to be in a definitely better position than quantum gravity.  This is the issue of unification, where loop gravity has nothing to offer. 

I think that it is important to emphasize the fact that the unification of the forces and the quantization of gravity are two conceptually distinct problems. The first is the old dream of having a single theory explaining everything.  The second refers specifically to the present inconsistency between general relativity and the standard formulation of quantum field theory, and is a problem that has to be solved in order to have a coherent theory of the world.  Solving the second does not necessarily imply solving the first: the quantum theory of the gravitational field can in principle be found without addressing the unification problem, like the quantum theory of the strong interactions has been understood and found without solving the problem of unifying them with the electroweak forces. 

The idea is often put forward that the problem of quantum gravity can \emph{only} be solved together with the unification problem. There are  hints that this might be the case. The running of the Standard Model coupling constants appears to converge not too far from Planck scale, fermions and boson divergences tend to cancel in supersymmetric theories, and so on. In the history of physics, often two major problems have been solved at once, and the temptation to do the trick again is reasonable.  

But even more often, however, hopes to solve two problems at once have been disappointed.  When I was a student, the idea that the theory for the strong interactions could only be found by getting rid of renormalization theory at the same time, was an unquestioned mantra, repeated by everybody.  It turned out to be wrong. There are standard arguments against the possibility of finding a consistent quantum theory of general relativity alone.  But these arguments hold in the context of standard \emph{local} field theory, where fields operators are defined on a spacetime metric  manifold. They are all circumvented by loop gravity by moving up to the proper context of a  \emph{general-covariant} quantum field theory.

Loop quantum gravity is a theory of quantum gravity that does not address the unification problem. It is like QED, or, more precisely, QCD: a quantum field theory for a certain interaction, which can be coupled to other interactions (affecting them), but is consistent by itself.  The philosophy underlying loop gravity is that we are not near the end of physics, we better not dream of a final theory of everything, and we better solve one problem at the time, which is hard enough.  

Back to the unification problem, does string theory actually solve it? Closed and open strings describe gravity and gauge theory.  More than that, they can even be shown as two sides of the same physics, under certain conditions.  This is very compelling.  String theory definitely provides a unified picture in which gauge theory and gravity live together, and the nineteen or so parameters of the Standard Model are replaced by a single fundamental parameter.  This is a strong plus.

But the initial objective of unification was far more ambitious.  It was to understand what is beyond the Standard Model in order to be able to \emph{compute} the value of the free parameters of the theory, in the same manner in which the Shr\"odinger equation allows us to compute chemical or condensed-matter parameters from fundamental constants.  There is no computation of the Standard Model parameters from string theory.  Nor a solution to the other puzzles in the theory: why is the cosmological constant so small? What is the origin of the three families? Can we give a better account of symmetry breaking?   Little concrete physics has emerged from the theory so far. The  results expected from a true unification do not seem to me to be there. 

\section{Applications} 

Black holes thermodynamics is definitely a success of string theory, and in my opinion, the strongest evidence for its physical relevance.   A similar success can be claimed by loop gravity.  Both successes are partial in my opinion.  The string derivation is still confined to, or around, extreme situations, as far as I know, and since it is based on mapping the physical black-hole solution into a different solution, it fails to give us a direct-hand concrete understanding of the relevant black hole degrees of freedom, as far as I can see.  The loop derivation of black hole entropy gives a clear and compelling physical picture of the relevant degrees of freedom contributing to the entropy, but it is based on tuning a free parameter to get the correct Bekenstein-Hawking entropy coefficients, and this does not sound satisfactory to me either. 

The crucial application to both strings and loops will probably turn out to be cosmology.  This is the most likely domain where a window of opportunity for testing the theories might open. Loop cosmology is the most spectacular success of loop quantum gravity.  The theory elegantly resolves the big bang singularity and predicts a sort of ``bounce" from a previously contracting phase.  When a  collapsing universe reaches Planck-scale density its wave function opens up into a genuinely quantum state where classical space and time are ill defined.  The quantum equation of the theory continue to hold, and the evolution can be studied across this non-classical region into a new expanding universe.  

This is similar of the picture of an electron falling straight into a Coulomb potential: the classical trajectory falls into the singularity. But the classical trajectory becomes ill-defined in the quantum evolution of the corresponding wave packet.  Spacetime is ill-defined around the big bang like the classical trajectory of the electron around the center of the Coulomb potential.   

The full quantum gravity effects are nicely summarized into an effective Planck-scale term that modifies the Friedmann equation, and an effort is under way to explore eventual testable consequences. Furthermore, inflation appear to be generic in this picture.  The picture is simple, physically compelling and based only on standard general relativity and quantum mechanics, empirically well established physical inputs. 

String cosmology is much developing as well, in a number of variants. The ability of the string to effectively resolve singularities and the possibility of topology change potentially provides important inputs to cosmology.   I might be wrong, and this is vague, but for my general relativist formation, however, many concrete scenarios proposed by string theory to describe the big bang, in particular some brane cosmologies with configurations of branes in a background space-time, do not sound physically very plausible to me, compared to the clean simplicity of the loop-cosmology scenario.  

Finally, string theory techniques may have potential applications to other domains of physics. These are  very interesting, but in no way they testify in favor of the relevance of string theory for the fundamental interactions.  Enormous intellectual investments have gone into string theory in the last decades and it would be strange if all the theoretical technology developed did not turn out to be good for something.  Theoretical physics is pretty coherent and techniques developed in one field often turn out to be helpful elsewhere, irrespective of their success in the first place.  After all, if string theory turned out to be useful for QCD, it would, in a sense, finally fulfill the aim for which its ideas were conceived at its very early initial stage, when Gabriele Veneziano wrote the dual amplitudes to describe strong interactions.  To some approximation, there certainly are  strings in the real world: the flux tubes of a confining gauge  interaction.

\section{Predictions}

Finally, although this should have probably been the first section, the main shortcoming of string theory is definitely its failure, so far, to produce any concretely verifiable physical prediction.   To be sure, string theory has provided numerous ``predictions", like short scale modifications of the gravitational force, black holes at CERN, dielectron resonances, or the existence of super-symmetric particles at low energy, but so far all these ``predictions" have been falsified by observation.  The theory has survived these failed predictions, because they were not solid predictions, but only hints of possibilities, effects compatible with the theory, but not necessary consequences of the theory.  The real problem is that the theory does not appear, so far, to have any verifiable necessary consequence at accessible scales. 
 
A burning difficulties is of course the landscape problem. If there is an accurate string description of the real world, then there are probably so many of them to make the discovery of the right one virtually impossible and in any case devoid of predictive power. 

In my opinion, this is serious.  A physical theory that does not give predictions is not a good theory.  We need definite predictions, like those that \emph{all} good physical theories of the past have been able to produce.  

Sometimes the strategy of saying ``so is the world, we have to live with this", is put forward.  I find this strategy unconvincing.  Such a strategy would be questionable even if string theory had already proved itself as a physically correct theory of the world.   But concluding that fundamental physics cannot anymore make definite predictions, just because a hypothetical theory turns out to be too weak to be predictive, is mistaking hypotheses for consequences. 

As far as clear verifiable predictions are concerned, loop quantum gravity is in no better shape either. There are no experiments supporting loops, nor any other quantum theory of gravity.  The simple question I have emphasized in the introduction --what is the scattering amplitude for two particles interacting gravitationally with a center of mass energy of the order of the impact parameter in Planck units?-- does not have a clear answer yet, neither from strings nor from loops. Therefore the above condemnation of string theory applies equally to all other approaches to the problem of quantum gravity.   

The closest to a verified prediction in the domain, as far as I know, comes from the poset approach to quantum gravity, which indicated the correct order of magnitude of the cosmological constant before its measurement \cite{Ahmed:2002mj}.  In this particular regard, string theory features particularly badly: not only it failed to predict a positive cosmological constant, but the very introduction of a positive cosmological constant appears to be at least problematic for the theory.

\section{``It does not work, therefore let's develop it further"}

I think that the problem of describing our physical world at the elementary level beyond current established theories is open.  String theory is one of the research directions among others aiming at solving this problem, with points of strength and weakness.  Its main strength is its mathematical construction where gauge fields, fermions and the gravitational field can be seen as parts of an overall coherent construct.  The theory has not delivered what it seemed to be almost there twenty years ago: a finite theory where the fundamental degrees of freedom are clearly identified, capable of describing our own world, with three fermion families, the $SU(3)\times SU(2) \times U(1)$ gauge groups, the values of parameters of the standard model computable, and (we should add today) a small positive cosmological constant. The tentative predictions of the theory have so far been falsified.  The development of the theory has constructed a toolbox that can perhaps be used in other contexts, but for the moment does not appear very effective for producing concrete results for high-energy physics.  The picture of quantum geometry offered by the theory is still very unclear. 

There has been a tremendous theoretical investment on strings, by far unmatched by alternative research directions, there have been successes, string revolutions and excitement. Seven years ago, I wrote a playful ``dialog" to point out what I saw as the theory's shortcomings at the time \cite{Rovelli:2003wd}; reading the dialog today, it seems to me that those same difficulties are still open.  Contrary to this, a theory like loop gravity has developed because the key problems open seven years ago have since been solved. 

There is a compelling logical evolution that has lead from the particle-theory successes to strings.  This path has been characterized by a sequence of spectacularly successful predictions (antiparticles, neutral currents, $W$ and $Z$, various quarks, just to mention some) which at some point has turned into a sequence of spectacularly failed predictions (grand unified theories predicted proton-decay at $10^{31}$ years,   Kaluza-Klein theory predicted an observable scalar field, strings suggested effects of extra dimensions,  supersymmetry has been ``on the verge of being seen" year after year, \ldots) I think that we should keep in mind the possibility that a wrong turn might have been taken at some point along this path. 

In recent years, various theories have developed following the logic: ``it does not work, therefore let's develop it further". Perseverance may pay (it worked with Yang-Mills theories), but at a risk: a theory can grow on its own failures, enriching its structures to cover previous insuccesses. 

There is certainly much beauty in strings. But beautiful ideas have turned out to be wrong in science, even ideas developed by large groups of scientists.  (In the words of a quote attributed to Thomas Henry Huxley: ``Science is organized common sense where many a beautiful theory was killed by an ugly fact".) The history of quantum gravity is particularly sprinkled with great hopes disappointed.  I remember as a young student sitting in a major conference where a world-renewed physicist announced that the definitive theory of quantum-gravity-and-everything, had finally been found.  There were a few skeptics in the audience, regarded as zombies by the majority. Today most of us do not even remember the name of that ``final theory". Or worse, we can think of more than one possibility \ldots  

It is obviously \emph{not} my intention to suggest that research in string theory should not be vigorously pursued. 
String theory is a spectacular intellectual achievement and it might well turn out to be the right track.  It is a rich and elaborate theory, that deserves to be studied further, with the resolute aim of arriving at assessing its physical viability.  If all the hopes of the string community are realized, it is a triumph. 

But I think it would be a mistake to consider string theory as an established result about nature and therefore concentrate the attention solely on it.  Also if the hopes of other research directions are realized, it would be a triumph.  String theory appears of unmatched beauty to string theorists, but other ideas appear of unmatched beauty to others.   What I think is important is to keep in mind that these theories are provisional. 

I am not pessimistic. Major problems like the ones we are facing have sometimes resisted for a while in the history of physics, but a solution has generally been found eventually.   The issue is open. I think that different path must be pursued.  Completeness, internal consistency, full agreement with known low-energy physics, simplicity, and, ultimately, experience, will tell. 

\vskip3mm
\centerline{---}
\vskip1mm

I thank two anonymous referees for very useful criticisms and Matthias Blau for a detailed reading of the paper and a useful conversation.   Thanks to Sal, a bit grown up and still without a job, for comments and suggestions.

\end{document}